\documentclass[prd,reprint,amsmath,amssymb,preprintnumbers,superscriptaddress,nofootinbib]{revtex4-1}
\usepackage[colorlinks,dvipdfmx]{hyperref}
\usepackage[dvipdfmx]{color}
\usepackage[dvipdfmx]{graphicx}

\begin{document}


\preprint{CTPU-17-11}
\title{
\Large
Hillclimbing inflation
}
\renewcommand{\thefootnote}{\alph{footnote}}

\author{
Ryusuke Jinno
}
\author{
Kunio Kaneta
}

\affiliation{
Center for Theoretical Physics of the Universe, Institute for Basic Science (IBS), Daejeon 34051, Korea
}

\begin{abstract}
We propose a new class of inflationary models 
in which inflation takes place while the inflaton is climbing up a potential hill due to a coupling to gravity.
We study their attractor behavior, and investigate its relation with known attractors.
We also discuss a possible realization of this type of models with the natural inflation, 
and show that the inflationary predictions come well within the region consistent with the observation of the cosmic microwave background.
\end{abstract}

\maketitle

\section{Introduction}

Cosmological inflation plays an essential role in addressing 
various cosmological issues~\cite{Guth:1980zm} as well as 
generating the primordial perturbations~\cite{Mukhanov:1981xt}.
Although the idea of inflation leads to a successful picture in modern cosmology, 
the underlying particle physics is still unclear.
Among various possibilities, one of the most attractive scenarios is extending the gravity sector.
The Starobinsky inflation~\cite{Starobinsky:1980te} is one of the most successful models along this line,
and the Higgs inflation~\cite{Lucchin:1985ip,Futamase:1987ua,Salopek:1988qh,CervantesCota:1995tz,Bezrukov:2007ep}, in which a nonminimal coupling between the Higgs field and the Ricci scalar is introduced,
has also persisted to date.

Over the past few years, our understanding on the behavior of these inflation models 
has been significantly developed: the discovery of attractor. 
It has been found that a large class of models with a general form of nonminimal coupling to gravity, 
including the Higgs inflation, has similar inflationary predictions.
Generalization of such models has been made (``universal attractor"~\cite{Kallosh:2013tua}
or ``induced inflation"~\cite{Giudice:2014toa,Kallosh:2014laa}),
and their attractor behavior is now called ``$\xi$-attractor"~\cite{Galante:2014ifa}.
On the other hand, the recently proposed ``$\alpha$-attractor"~\cite{Kallosh:2013yoa} has revealed
a generic feature of attractors appearing in the models with a kinetic pole,
which coincide with $\xi$-attractor with a certain choice of model parameters~\cite{Galante:2014ifa}.

In this paper, we revisit inflation models with the nonminimal coupling.
We first propose a new class of models which is featured by the climbing of the inflaton up the potential hill
by considering a specific behavior of the nonminimal coupling.
We point out that this class of models has attractor behavior, 
and discuss its relation with known attractors in a general way.
Then we give a concrete example of such models 
using the natural inflation-type potential~\cite{Freese:1990rb} with a nonminimal coupling,
and show that the inflationary predictions come well within the region consistent 
with Planck observation of the cosmic microwave background (CMB)~\cite{Ade:2015xua}.
We also point out that our setup can be applied to a broad class of inflaton potentials 
which have multiple degenerate vacua.

The organization of the paper is as follows.
In Sec.~\ref{sec:HillclimbingGeneral} 
we give general discussion on the attractor behavior in inflation models with the nonminimal coupling.
In Sec.~\ref{sec:HillclimbingExample} we give a concrete model to illustrate the point discussed
in Sec.~\ref{sec:HillclimbingGeneral} and show the inflationary predictions.
We finally conclude in Sec.~\ref{sec:Conclusion}.

\section{Attractors and hillclimbing inflation}
\label{sec:HillclimbingGeneral}

\subsection*{$\xi$-attractor}

Let us start with general discussion on attractor solutions in inflation models with the nonminimal coupling.
Throughout the paper, we focus on the following single-field inflation setup:
\begin{align}
S
&=
\int d^4x \sqrt{-g_J} 
\left[
\frac{M_P^2}{2} \Omega R_J
- \frac{K_J}{2} (\partial_J \phi_J)^2 - V_J
\right],
\label{eq:SJ}
\end{align}
where $M_P$ is the reduced Planck mass,
and $\phi_J$, $R_J$ and $V_J(\phi_J)$ are 
the (Jordan-frame) inflaton, Ricci scalar and potential, respectively.
Hereafter the subscript $J$ indicates the ``Jordan frame''.
Also, the factor $K_J(\phi_J)$ in front of the kinetic term 
$(\partial_J \phi_J)^2 \equiv g^{\mu \nu}_J \partial_\mu \phi_J \partial_\nu \phi_J$
is an arbitrary function of $\phi_J$, which we retain for generality of the following argument.
In addition, $\Omega(\phi_J)$ is an arbitrary function which takes positive values
in the case of our interest. 
Under the Weyl rescaling $g_{\mu \nu} \equiv \Omega g_{J \mu \nu}$,
the Ricci scalar transforms as
\begin{align}
R_J
&=
\Omega 
\left[ 
R + 3 \Box \ln \Omega 
- \frac{3}{2} (\partial \ln \Omega)^2
\right]
\label{eq:R}
\end{align}
with which the action (\ref{eq:SJ}) becomes
\begin{align}
S
&=
\int d^4x \sqrt{-g} 
\left[
\frac{M_P^2}{2}R
- \frac{K}{2} (\partial \phi_J)^2 - V
\right],
\label{eq:SE}
\end{align}
where the potential is given by $V = V_J / \Omega^2$, 
and
\begin{align}
K
&= 
\frac{K_J}{\Omega}
+
\frac{3M_P^2}{2\Omega^2}
\left(
	\frac{d\Omega}{d\phi_J}
\right)^2.
\label{eq:KE}
\end{align}
Here let us suppose that the second term dominates the first term in Eq.~(\ref{eq:KE}).
Such a setup occurs {\it e.g.}
when $M_P^2 (d\Omega/d\phi_J)^2 / \Omega^2 \gg 1/\Omega$ for $K_J = 1$
or when $K_J = 0$.
Then the kinetic term in the action (\ref{eq:SE}) reduces to~\cite{Kallosh:2013tua}
\begin{align}
&
-\frac{K}{2}(\partial \phi_J)^2 \simeq - \frac{3M_P^2}{4} (\partial \ln \Omega)^2.
\label{eq:kinE}
\end{align}
Now let us assume that 
$\Omega$ evolves from $\Omega \gg 1$ to $\Omega = 1$ during inflation.\footnote{
One can take $\Omega = 1$ in the present universe without loss of generality.
}
For example, if we take $V = V_0(1 - \Omega^{-1})^2$ 
with $V_0$ being a constant which determines the potential height at $\Omega \gg 1$,
this potential realizes a plateau for $\Omega \gg 1$ 
and a vanishing cosmological constant in the present universe.\footnote{
As we see in the next subsection and show in footnote~\ref{fn:xiExpand},
this potential form can be generalized to a broader class.}
In terms of the Einstein-frame inflaton, one may identify 
$\phi / M_P \simeq \sqrt{3/2} \ln \Omega$
to make the kinetic term canonical, and then the potential becomes
\begin{align}
V
&= 
V_0 
\left( 1 - e^{- \sqrt{\frac{2}{3}} \frac{\phi}{M_P}} \right)^2.
\label{eq:VEzeta}
\end{align}
This potential realizes the spectral index $n_s$ and the tensor-to-scalar ratio $r$,
\begin{align}
n_s 
&\simeq 1 - \frac{2}{N},
\;\;\;\;
r 
\simeq \frac{12}{N^2},
\label{eq:nsrxi}
\end{align}
in large-$N$ limit with $N$ being the $e$-folding number.
It is known that a large class of inflation models predicts Eq.~(\ref{eq:nsrxi}), 
which includes the Higgs inflation 
$\Omega = 1 + \xi \phi_J^2/M_P^2$~\cite{Futamase:1987ua,CervantesCota:1995tz,Bezrukov:2007ep} 
and its generalizations such as
``universal attractor" $\Omega = 1 + \xi f(\phi_J)$~\cite{Kallosh:2013tua}
or
``induced inflation" $\Omega = \xi f(\phi_J)$~\cite{Giudice:2014toa,Kallosh:2014laa}.
In Ref.~\cite{Galante:2014ifa},
the class of models which have this attractor behavior in the inflationary predictions (\ref{eq:nsrxi})
have been dubbed as ``$\xi$-attractor".

\subsection*{$\eta$-attractor}

Here we take a closer look at the arguments in the previous subsection.
In discussing $\xi$-attractor, we have assumed that the inflationary regime occurs at $\Omega \gg 1$.
In this paper, in contrast, we consider
\renewcommand{\labelitemi}{}
\begin{itemize}
\item
\begin{center}
Inflation at $\Omega \ll 1$.
\end{center}
\end{itemize}
This distinction is important from a model-building point of view,
as we see in the rest of this subsection and also in the next section with a concrete example.

To investigate the properties of such inflation models,
let us first consider the requirements to realize a vanishing cosmological constant 
in the present universe.
Assuming that $\Omega$ is monotonic for the inflaton values of our interest,
we write down the potential in terms of $\Omega$:
\begin{align}
V
&= 
V_0 
\left( 1 - \sum_{k = n}^\infty \eta_k \Omega^k \right),
\label{eq:VEeta}
\end{align}
with $\eta_k$ being some constants and $n \geq 1$ being the leading exponent of $\Omega$
which dominantly affects the inflationary predictions.
Note that we have not included negative powers of $\Omega$,
in order to keep the flatness of the potential.
The vanishing cosmological constant is realized by
\begin{align}
\sum_{k = n}^\infty \eta_k
&= 1.
\end{align}
When the approximation (\ref{eq:kinE}) holds,\footnote{
See Appendix~\ref{app:Cond} for the cases in which this approximation does not hold.}
the slow-roll parameters $\epsilon$ and $\eta$ are given by
\begin{eqnarray}
	\epsilon &\equiv& \frac{M_P^2}{2}
	\left(
		\frac{V'}{V}
	\right)^2
	\simeq \frac{1}{3}(n\eta_n\Omega^n)^2,
	\label{eq:eps} \\
	\eta &\equiv& M_P^2 \frac{V''}{V}
	\simeq -\frac{2}{3} n^2 \eta_n \Omega^n,
	\label{eq:eta}
\end{eqnarray}
where we respectively define $V'$ and $V''$ as $dV/d\phi$ and $d^2V/d\phi^2$, 
and thus the inflaton rolls slowly when $\Omega$ becomes sufficiently small. 
The attractor behavior of the inflationary predictions with the potential (\ref{eq:VEeta})
can be calculated as
\begin{align}
n_s
&\simeq 1 - \frac{2}{N},
\;\;\;\;
r
\simeq \frac{12}{n^2N^2},
\label{eq:nsreta}
\end{align}
in large-$N$ limit.
It should be noted that the resultant $r$ is more general than Eq.~(\ref{eq:nsrxi}).
Of course, in $\xi$-attractor as well,
the inflationary predictions reduce to Eq.~(\ref{eq:nsreta}) 
by adopting a similar expansion to the potential.\footnote{
\label{fn:xiExpand}
The same line of argument is possible for $\Omega \gg 1$ as well.
One may write down the potential as
\begin{align}
V_E 
&= 
V_0 
\left( 1 - \sum_{k = n}^\infty \xi_k \Omega^{-k} \right),
\label{eq:VExi}
\end{align}
with $\xi_k$ being some constants and $n$ being the leading exponent.
Here we have not included positive powers of $\Omega$
because such terms spoil the flatness of the potential for $\Omega \gg 1$.
The vanishing cosmological constant in the present universe is realized by
\begin{align}
\sum_{k = n}^\infty \xi_k
&= 1.
\end{align}
Note that $V_E \sim (1 - \Omega^{-1})^2$ gives one realization of this condition.
The setup (\ref{eq:VExi}) gives the same prediction as Eq.~(\ref{eq:nsreta}),
which is more general than Eq.~(\ref{eq:nsrxi}).}\footnote{
In Ref.~\cite{Yi:2016jqr}, generalization of $\xi$-attractor has been made
by setting $V_E \sim (1 - \Omega^{-p})^2$ where $p$ can be other than unity.
}
However, as we see in Sec.~\ref{sec:HillclimbingExample}, 
relatively simple setups lead to $n \neq 1$ in this type of inflation models, 
and therefore we call them “$\eta$-attractor” in the following to distinguish 
between the two.\footnote{
In both of the two classes,
inflationary predictions generically deviate from the attractor limit (\ref{eq:nsreta})
once one considers concrete model constructions.
Therefore it would be possible to distinguish such models by future CMB observations.}

There is another important reason to distinguish between $\xi$- and $\eta$-attractors:
application to existing inflaton potentials.
To see this, let us consider the inflaton behavior in the Jordan frame.
The Jordan-frame potential is given by 
$V_J = \Omega^2 V_E \simeq \Omega^2 V_0$ during inflation.
This means that, in the class of inflation models which is featured by $\Omega \ll 1$, 
the inflaton {\it climbs up the potential hill} due to a coupling to gravity.
If one considers applying such a gravity effect to existing inflaton potentials,
one sees that the existence of $\eta$-attractor can help the predictions of 
inflation models with multiple potential minima by modifying the inflaton behavior.
We call such inflation models with $\eta$-attractor behavior 
{\it hillclimbing inflation} in the rest of the paper.
We will illustrate this point in Sec.~\ref{sec:HillclimbingExample} with a concrete example.

Before moving on to the next topic, let us mention previous studies. 
It is known that in some specific setup the inflation can take place with a small $\Omega$, 
for instance, the conformal inflation 
with $\chi=\sqrt{6}$ gauge~\cite{Kallosh:2013hoa,Kallosh:2013daa}
and its extension~\cite{Kallosh:2013maa}. 
While they are based on a specific type of inflation models, 
we stress that this class of inflation can be seen in a rather broader range of models 
having multiple vacua in the inflaton potential, as we see in the next section.

\subsection*{Relation with $\alpha$-attractor}

\begin{figure}
\begin{center}
\includegraphics[width=\columnwidth]{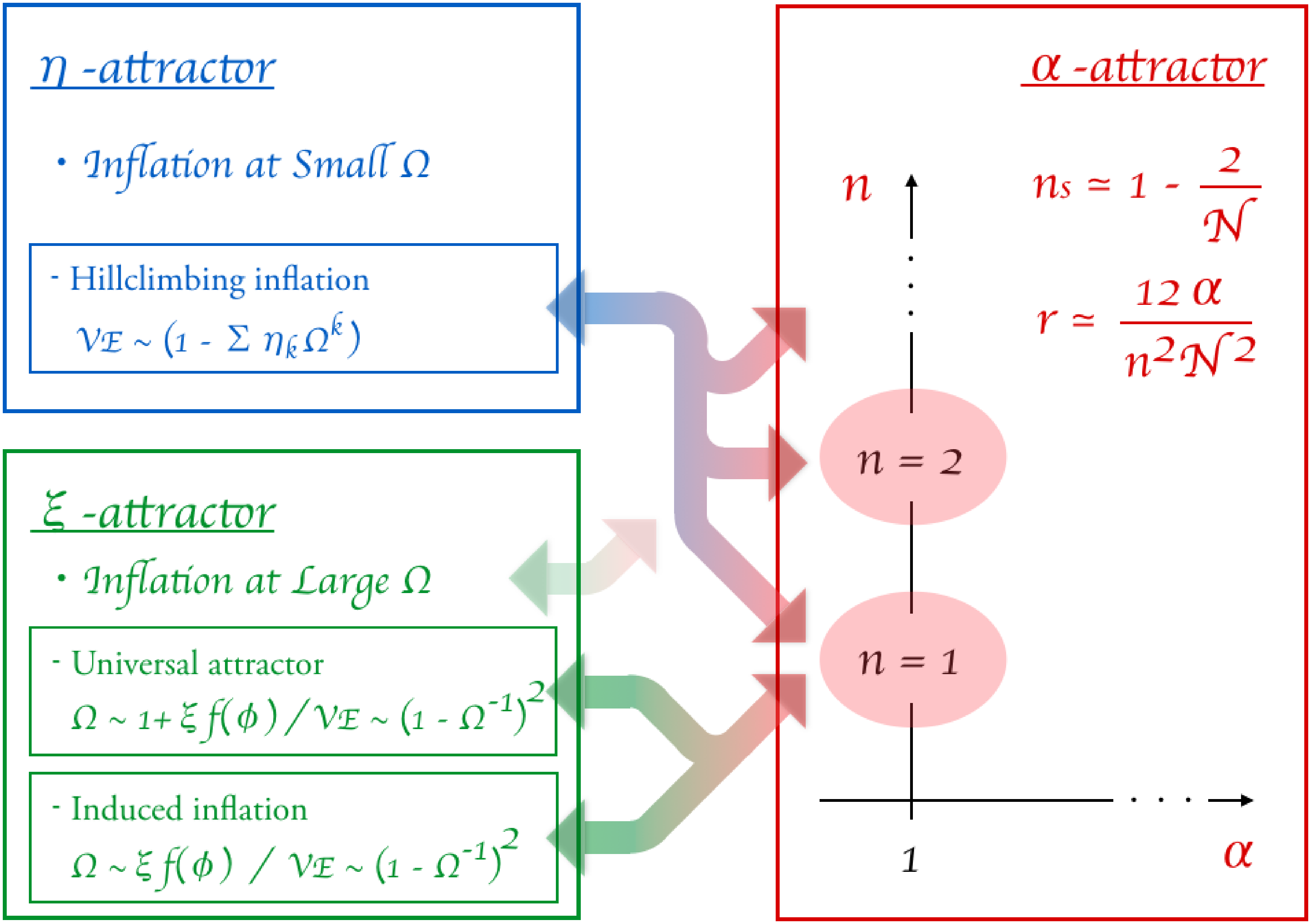}
\caption {\small
Relation between $\xi$- or $\eta$-attractor and $\alpha$-attractor.
For both $\xi$- and $\eta$-attractors, the corresponding value of $\alpha$ is unity 
because it is determined by the residue of the kinetic pole.
Well-known examples of $\xi$-attractor such as Higgs inflation and its generalizations
correspond to $n = 1$ coming from $V \sim (1 - \Omega^{-1})^2 = 1 - 2\Omega^{-1} + \cdots$.
For $\eta$-attractor, $n > 1$ can easily be realized as we see in Sec.~\ref{sec:HillclimbingExample}. 
It is also possible to generalize $\xi$-attractor to more general values of $n$
as mentioned in footnote~\ref{fn:xiExpand}.}
\label{fig:attractor}
\end{center}
\end{figure}

Here we comment on the relation with $\alpha$-attractor, 
which was discovered in the studies of 
superconformal inflation~\cite{Kallosh:2013wya,Kallosh:2013xya}
and later developed to the current 
form~\cite{Kallosh:2013hoa,Kallosh:2013daa,Kallosh:2013yoa,Kallosh:2014rga}.
Its action is given by~\cite{Kallosh:2013yoa,Kallosh:2014rga}
\begin{align}
S
&=
\int d^4x \sqrt{-g} \; {\mathcal L}, 
\label{eq:Salpha}
\\
{\mathcal L}
&= 
\frac{M_P^2}{2}R 
- \frac{\alpha}{\left(1 - \phi^2/6M_P^2\right)^2}(\partial \phi)^2 
- V \left( \phi/\sqrt{6}M_P \right),
\label{eq:Lalpha}
\end{align}
with non-negative potential $V$.\footnote{
The pole structure in the inflaton kinetic term
has been generalized in {\it e.g.} Ref.~\cite{Terada:2016nqg}.
}
This class of models coincides with the Starobinsky model
with $\alpha = 1$ and a specific choice of the potential~\cite{Kallosh:2013wya}.
Canonical normalization of the inflation field 
$\phi/\sqrt{6}M_P = \tanh \varphi/\sqrt{6\alpha}M_P$
leads to
\begin{align}
{\mathcal L}
&=
\frac{M_P^2}{2}R - \frac{1}{2}(\partial \varphi)^2 - V \left( \tanh \frac{\varphi}{\sqrt{6\alpha}M_P} \right).
\end{align}
The potential can generically be expanded as 
\begin{align}
V
&=
V_0
\left(
1 - \sum_{k = n} \alpha_k e^{- k \sqrt{\frac{2}{3\alpha}} \frac{\varphi}{M_P}} 
\right),
\label{eq:VEexpandalpha}
\end{align}
with $\alpha_k$ being some constants and 
$n$ being the leading exponent as in the previous subsections.
Though usually $n = 1$ is assumed in calculating inflationary predictions in this class of 
models~\cite{Kallosh:2013hoa,Kallosh:2013daa,Kallosh:2013yoa,Kallosh:2014rga,Galante:2014ifa},
which holds true for {\it e.g.} 
$V \sim (1 -  e^{- \sqrt{\frac{2}{3\alpha}} \frac{\varphi}{M_P}})^2$,
we do not restrict ourselves to such a special case.
The inflationary predictions for the potential (\ref{eq:VEexpandalpha}) approach
\begin{align}
n_s
&\simeq 1 - \frac{2}{N},
\;\;\;\;
r
\simeq \frac{12\alpha}{n^2N^2},
\label{eq:nsralpha}
\end{align}
in large-$N$ limit.
The correspondence between $\xi$- or $\eta$-attractors 
and $\alpha$-attractor is easily seen if one notice that 
the pole structure and the leading exponent dominantly determine 
the inflationary predictions~\cite{Galante:2014ifa}:
$\alpha$-attractor, 
whose action is given by Eqs.~(\ref{eq:Salpha})--(\ref{eq:Lalpha}),
shares its inflationary predictions for $\alpha = 1$
with those of $\xi$- or $\eta$-attractors, whose action is given by Eq.~(\ref{eq:SJ})
and inflation occurs at $\Omega \gg 1$ and $\Omega \ll 1$, respectively.
We summarize the relation in Fig.~\ref{fig:attractor}.

\section{Hillclimbing natural inflation}
\label{sec:HillclimbingExample}

\begin{figure}
\begin{center}
\includegraphics[width=0.9\columnwidth]{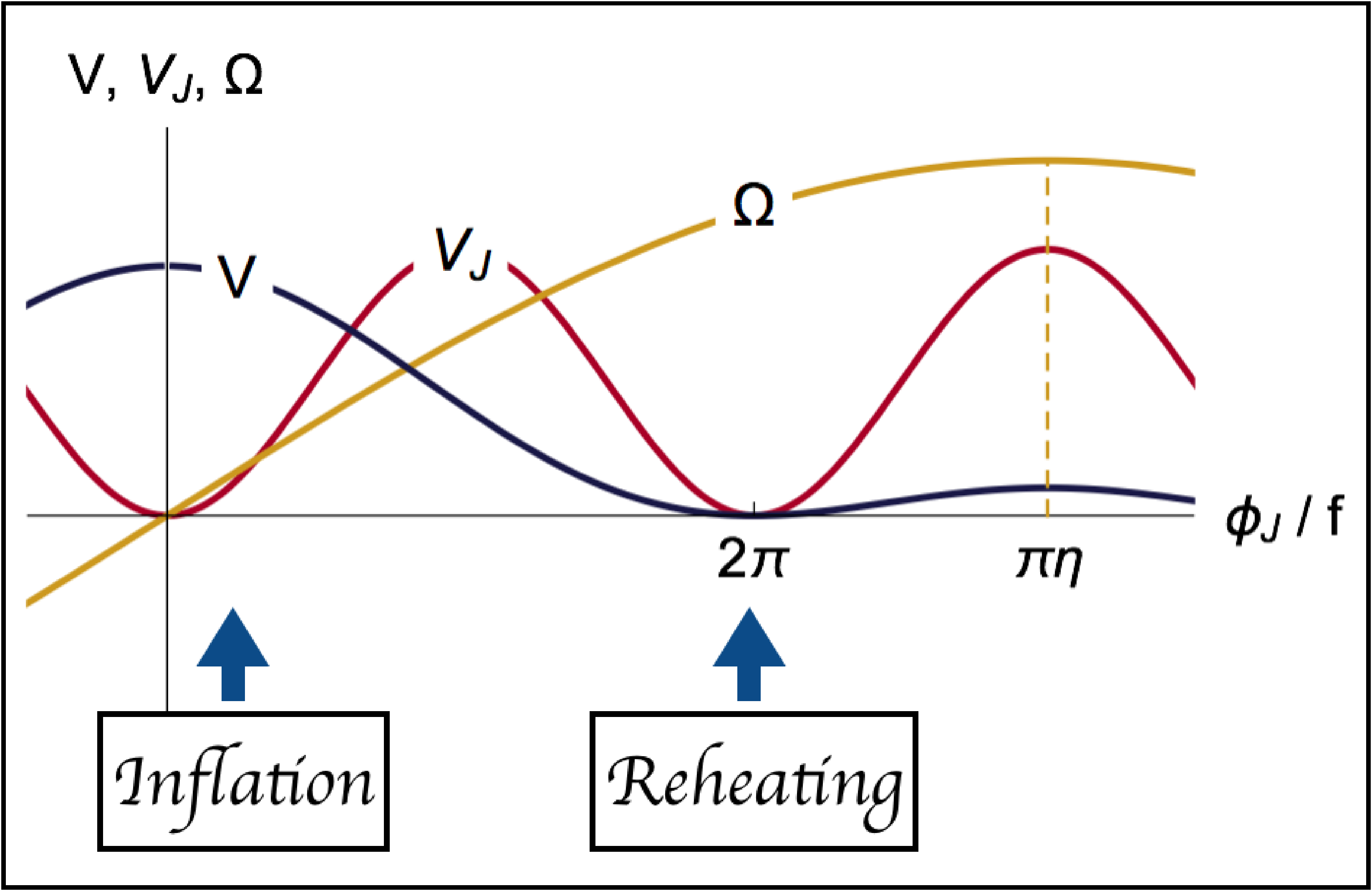}
\caption {\small
Rough sketch of our setup.
The blue, red and yellow lines correspond to
the Einstein-frame potential $V$, 
the Jordan-frame potential $V_J$ 
and the conformal factor $\Omega$, respectively.
In this setup, inflation occurs around the origin 
while the inflaton is climbing up the Jordan-frame potential,
and ends at the neighboring minimum.
}
\label{fig:Pot}
\end{center}
\end{figure}

Now let us illustrate our point discussed in the previous section with a specific example,
which we call {\it hillclimbing natural inflation}.
The model setup is the Jordan-frame action (\ref{eq:SJ}) with $K_J = 1$ and
\begin{align}
\Omega
&=
\omega \sin \left(  
\frac{\phi_J}{2 \eta f} 
\right),
\label{eq:HCNOmega}
\\
V_J
&=
\Lambda^4
\left[
1 - \cos \left( \frac{\phi_J}{f} \right)
\right].
\label{eq:HCNVJ}
\end{align}
Here $\omega$, $\eta$, $f$ and $\Lambda$ are free parameters of the model (see also Appendix \ref{app:Models}).
We set the coefficient $\omega$ to satisfy
\begin{align}
\omega
&= 
1 \Big/ \sin \left( \frac{\pi}{\eta} \right),
\end{align}
to realize $\Omega =1$ in the present universe.
Note that other choices of $K_J$ such as $K_J = 0$ also work,
because they only give the negligible part in Eq.~(\ref{eq:KE}).
Figure~\ref{fig:Pot} shows typical shapes of the potential and $\Omega$ in this model.

In this setup the inflation takes place in the vicinity of the origin, 
and the inflaton is slowly displaced towards $\phi_J > 0$ by climbing up the potential hill of $V_J$ 
due to the small conformal factor.
This property can also be understood in the Einstein frame.
The relation between $\phi_J$ and $\phi$ is given by
\begin{align}
\phi_J \propto e^{\sqrt{\frac{2}{3}} \frac{\phi}{M_P}}
\end{align}
around the origin.
Inflation occurs at $\phi_J \to +0$, or $\phi \to -\infty$,
where the potential $V = V_J / \Omega^2$ approaches to a certain constant value,
and ends at the neighboring minimum.
As long as we take $f \ll M_P$, the relevant inflationary dynamics occurs
around the origin and therefore the predictions are given by Eq.~(\ref{eq:nsreta}).
It is also seen that this model corresponds to the $n = 2$ case in Sec.~\ref{sec:HillclimbingGeneral}:
Noting that the Einstein-frame potential $V = V_J / \Omega^2$ is even in $\phi_J$,
one sees that only even powers of $\Omega$, which is an odd function of $\phi_J$, 
appear in the expansion of $V$.
As a result $n = 2$ (more specifically $\eta_2=(2/3)(\eta/\omega)^2$) 
appears as the leading exponent in Eq.~(\ref{eq:VEeta}).\footnote{
Note that $n = 1$ can also be realized easily
by choosing $\Omega$ not to be odd in $\phi_J$.
}
On the other hand, in the opposite case $f \gg M_P$, 
the relevant inflation dynamics occurs around the minimum $\phi_J = 2 \pi f$.
In this case, the conformal factor is close to unity all the way from the CMB scale to the end of inflation, 
and thus the model approaches to the quadratic chaotic inflation (see Appendix \ref{app:Cond} for more detail).

In Fig.~\ref{fig:nsr} we show the inflationary predictions of the hillclimbing natural inflation.
We have calculated the scalar spectral index $n_s$ and the tensor-to-scalar ratio $r$,
varying $f$ from large $f$ ($\gg M_P$) to small $f$ ($\ll M_P$).
In this figure, we have fixed the scalar normalization by ${\mathcal P}_\zeta = 2 \times 10^{-9}$
and taken the $e$-folding to be $N = 50$ and $60$.
It is seen that for $f \gg M_P$ the predictions coincide with those of quadratic chaotic inflation,
while they approach asymptotic values for $f \ll M_P$, which corresponds to 
the $n=2$ case of the $\eta$-attractor discussed in Sec.~\ref{sec:HillclimbingGeneral}.

\begin{figure}
\begin{center}
\includegraphics[width=0.9\columnwidth]{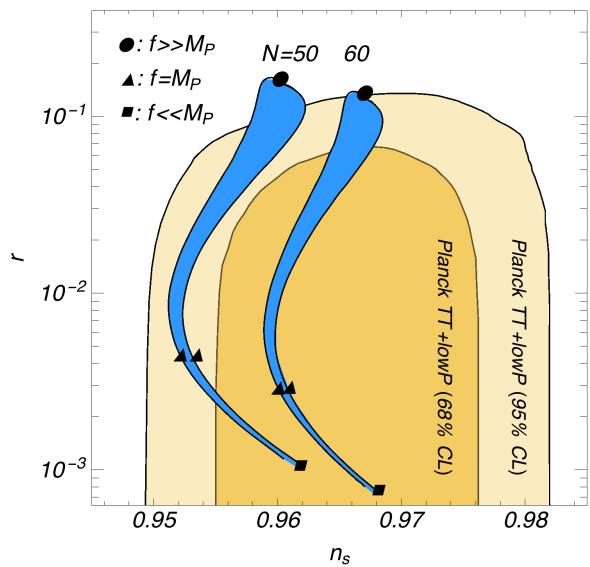}
\caption {\small
Predictions for inflationary parameters in the hillclimbing natural inflation.
In each of $N=50$ and $60$, 
we take $\eta=1.5$ to $4$ corresponding to the right and left boundaries, respectively.
Other choices of $\eta$, such as $\eta \gtrsim 4$, 
give almost the same curve as the one with $\eta = 4$.
}
\label{fig:nsr}
\end{center}
\end{figure}

\section{Conclusion}
\label{sec:Conclusion}

In this paper, 
we have presented a new class of inflationary models in which inflation takes place 
while the inflaton is climbing up a potential hill due to a coupling to gravity.
We have studied the attractor behavior in the resulting inflationary predictions 
and investigated its relation with known attractors, and as a result, we have proposed ``$\eta$-attractor".
The inflaton behavior in $\eta$-attractor is well understood 
in the Einstein frame, where the original potential is lifted up by the Weyl transformation.
We have also discussed a possible realization of this type of models with the natural inflation potential, 
and shown that the inflationary predictions are affected by the existence of the attractor.
Though in this paper we have restricted ourselves to an example with the natural inflation,
our discussion is also applicable to various types of models which have multiple vacua 
with vanishing vacuum energy.
For example, it would be interesting to investigate the possibility of realizing the hillclimbing inflation 
when the SM Higgs has another degenerate vacuum around $\sim 10^{17}$~GeV, as suggested by
the multiple-point principle~\cite{Bennett:1993pj}.
As another interesting possibility, the hillclimbing inflation would take place 
even if the inflaton potential in the Jordan frame is not bounded from below 
as long as the conformal factor becomes sufficiently small at a certain relevant scale.
We leave such studies as future work~\cite{JK}.

\section*{Acknowledgments}

This work was supported by IBS under the project code, IBS-R018-D1.
The authors are grateful to T.~Terada for helpful comments.

\appendix

\section{Models}
\label{app:Models}

In this appendix we discuss some realizations of the setup (\ref{eq:HCNOmega})--(\ref{eq:HCNVJ})
using a complex scalar field.
We consider the action $S = \int d^4x \sqrt{-g_J} {\mathcal L}_J$ with
\begin{align}
{\mathcal L}_J
&=
- iM(\Phi_J - \Phi_J^\dagger)R_J
- |\partial \Phi_J|^2 
- V_J,
\label{eq:LModel1}
\end{align}
or
\begin{align}
{\mathcal L}_J
&=
- i(\Phi_J^2 - \Phi_J^{\dagger 2})R_J
- |\partial \Phi_J|^2 
- V_J,
\label{eq:LModel2}
\end{align}
with $\Phi_J$ being a complex scalar and $M$ being a dimensionful parameter, 
and the potential is given by
\begin{align}
V_J(\Phi_J)
&= 
\lambda \left[ |\Phi_J|^p 
- \frac{1}{2}(\Phi_J^p + \Phi_J^{\dagger p}) \right] 
+ V_{\rm SB}(\Phi_J).
\label{eq:VJModel}
\end{align}
Here $p$ is some integer, and $V_{\rm SB}$ is introduced in order to fix the radial value of $\Phi_J$.
After $\Phi_J$ develops non-zero vacuum expectation value $\langle \Phi_J \rangle\neq 0$, 
the inflaton field $\phi$ residing in the phase of $\Phi_J$, 
$\Phi_J=\langle \Phi_J \rangle \exp(i \phi/\sqrt{2}\langle \Phi_J \rangle)$, 
acquires the potential of the form given in Eq.~(\ref{eq:HCNVJ}) by taking $\Lambda=\lambda \langle \Phi_J \rangle^p$ 
and $f=\sqrt{2}\langle \Phi_J \rangle/p$.\footnote{
In terms of shift symmetry along to the $\phi$ direction, 
the coupling $\lambda$ is regarded as an explicit breaking parameter. In the gravity sector, 
the nonminimal coupling also breaks the shift symmetry, 
and in some case the quantum corrections to the inflaton potential might affect the inflaton dynamics. 
In such a case, we may need a UV description of the nonminimal coupling to control the corrections.
}

In the hillclimbing natural inflation models, the reheating process is also a non-trivial issue, 
since the potential shape relevant for the reheating epoch highly depends on the choice of the conformal factor.
Also, it is known that in inflation with the nonminimal coupling 
the direction perpendicular to the inflaton can be violently produced 
at the onset of preheating~\cite{Ema:2016dny},
and it would be interesting to study whether this occurs in the present setup.

\section{Hillclimbing condition}
\label{app:Cond}

In this appendix we discuss conditions for a successful hillclimbing inflation,
focusing on special cases in which 
$V_J \propto \phi_J^2$ and $\Omega \propto \phi_J$ hold around the potential minimum 
(which we take to be near the origin).
In particular, we take a closer look at conditions 
with which the approximation given by Eq.~(\ref{eq:kinE}) is justified
as a consequence of the second term domination in the R.H.S. of Eq.~(\ref{eq:KE}).
In the following we take $K_J = 1$ for concreteness.

Let us parameterize the conformal factor around the potential minimum as
\begin{align}
\Omega
&\sim 
\frac{\phi_J}{M},
\end{align}
where $M$ is some dimensionful quantity.
Note that $\Omega = 1$ in the present Universe means that 
the inflaton value at the end of inflation is parameterized as $\phi_J \sim M$.
In the hillclimbing natural inflation it corresponds to $M \sim \eta f$.
The first and second terms in Eq.~(\ref{eq:KE}) are estimated as
\begin{align}
\frac{1}{\Omega}
&\sim \frac{M}{\phi_J},
\;\;\;\;
\frac{3M_P^2}{2\Omega^2} \left( \frac{d\Omega}{d\phi_J} \right)^2
\sim \frac{M_P^2}{\phi_J^2},
\label{eq:1vs2}
\end{align}
and therefore the second term dominates for $\phi_J \lesssim M_P^2 / M$.
Now let us consider the following two cases: 
\begin{align}
({\rm i}) \; M \gg M_P,
\;\;\;\;\;\;\;\;
({\rm ii})\; M \ll M_P.
\end{align}
For (i), the first term dominates the second term in Eq.~(\ref{eq:KE}) for the inflaton value from the CMB scale to the inflation end, and thus the discussions below Eqs.~(\ref{eq:eps}) and (\ref{eq:eta}) do not hold.
In such a case, the inflationary predictions deviate from those of the $\eta$-attractor.
Instead, recalling that the hillclimbing inflation needs another minimum in the potential
in order to terminate the inflationary regime 
(see the setup in Sec.~\ref{sec:HillclimbingExample}), 
one sees that the setup approaches to chaotic inflation.
This is because the whole inflaton excursion from the CMB scale to the inflation end 
falls within the vicinity of this reheating minimum,
which makes the evolution of the conformal factor $\Omega$ within this inflaton range negligible.
This is why the inflationary predictions in the hillclimbing natural inflation approach 
the ones in quadratic chaotic inflation for $f \gg M_P$.

On the other hand, for (ii), 
the second term dominates in Eq.~(\ref{eq:KE}) for the inflaton values of our interest.
As a consistency check, let us first assume the second term domination in Eq.~(\ref{eq:KE}).
Then, from the discussions in Sec.~\ref{sec:HillclimbingGeneral}, 
inflation occurs for $\Omega \ll 1$ or equivalently $\phi_J \ll M$.
Now plugging this back into Eq.~(\ref{eq:1vs2}) 
one sees that the second term indeed dominates for this inflaton range.



\end{document}